\documentclass[12pt]{revtex4-1}
\usepackage{graphicx}
\usepackage{amssymb} 
\usepackage{amsmath}    
\usepackage{latexsym}   
\usepackage{bm}
\usepackage{times}
\usepackage{centernot}
\usepackage{float}
\usepackage{color}
\usepackage{wrapfig}
\usepackage{mathtools} 

\begin{document}
\title{Clocks without ``Time'' in Entangled-State Experiments}

\author{F. Hadi Madjid}
\email{gailmadjid@comcast.net}
  \affiliation{82 Powers Road, Concord, MA 01742, USA.}
\author{John M. Myers}
\email{myers@seas.harvard.edu}
\affiliation{Harvard School of Engineering and Applied Sciences, Cambridge, MA 02138, USA.}

\begin{abstract}
Entangled states of light exhibit measurable correlations between light
detections at separated locations.  These correlations are exploited in
entangled-state quantum key distribution.  To do so involves setting up and
maintaining a rhythm of communication among clocks at separated
locations. Here, we try to disentangle our thinking about clocks as used in
actual experiments from theories of time, such as special relativity or
general relativity, which already differ between each other.  Special
relativity intertwines the concept of time with a particular definition of
the synchronization of clocks, which precludes synchronizing every clock to
every other clock. General relativity imposes additional barriers to
synchronization, barriers that invite seeking an alternative depending on
any global concept of time.  To this end, we focus on how clocks are actually
used in some experimental situations.  We show how working with clocks without worrying about time makes it possible to generalize some
designs for quantum key distribution and also clarifies the need for
alternatives to the special-relativistic definition of synchronization.
\end{abstract}

\maketitle

\section{Introduction}
When I am meeting a friend for lunch at noon, I think, conventionally, of my
watch as ``telling time''; however, for some purposes we prefer an
alternative theory of clocks based not on any theory of time but on relations
that link a reading of one clock at the transmission of a signal to a reading
of another clock at the reception of that signal.  This frees us to think of
each clock as coming with a faster-slower lever that its user can manipulate,
according to the user's purpose. Such a theory is reported in Ref.\cite{19MST}.
Here we extend this theory for application to situations in which agents,
typically called Alice and Bob, make use of entangled photon pairs.

As an arena in which to consider the use of clocks, experiments with pairs of
entangled photons present are especially interesting, because they involve
locations at which two or more agents make use of clocks while they operate
detectors and have occasion to communicate with one another across
propagation delays.  The situation of multiple detections raises vexed
questions of the interpretation of quantum mechanics.  The literature is too
extensive to review here, but we point to questions announced long ago but
still worth attention, arising in relating quantum mechanics to spacetime
\cite{AA81,AA84b}.  Although these questions are unlikely to be resolved any
time soon to the general satisfaction of quantum physicists \cite{zeilinger},
we have a particular contribution to offer.  In 2005 we proved, within the
mathematics of quantum theory, that no evidence expressed as probabilities of
outcomes can ever determine a unique explanation in terms of quantum states
and linear operators, so that any choice of an explanation within the
framework of quantum physics involves reaching beyond logic \cite{05aop}.
Hence, logically, the choice of an explanation is unpredictable, thereby
showing a drastic unpredictability in physics, above and beyond quantum
uncertainty.  Multiple theories are always logically possible, including
theories about clocks.

We consider situations in which {\it symbol-handling agents}, linked by
quantum as well as classical communications, make use of clocks, for example,
the Alice and Bob that appear in descriptions of quantum cryptography.  By
symbols, we have in mind letters of an alphabet, or, more basically, bit
strings. We think of an agent as acting in steps, one step after another.  An
agent has a memory and can communicate with other agents.  Because the agent
operates one step after another, the agent handles symbols sequentially.  An
agent's sequence of symbols can include symbols written by the agent as well
as classical bits received from other agents, and also, in the quantum
context, symbols transmitted by other agents, reporting detections of photons.

A general purpose for the use of clocks is to establish relations between
symbols transmitted and symbols received.  More exactly, agents in
communication with one another use clocks to relate symbols possessed by one
agent with symbols possessed by another agent.  For this purpose, the clocks
have to be, in one sense or another, synchronized. Some other investigators
have addressed synchronization in conjunction with the employment of
entangled photon pairs. In the next two sections we review a few of their
reports \cite{00tittel,06k,09k,18k} to show how our point of view of focusing
on clocks without worrying about time widens the potential applicability of
some designs for quantum key distribution and clarifies the need for
alternatives to the special-relativistic definition of synchronization.

If we work with a notion of a global time coordinate, we base our thinking on
an assumption, whether that of classical physics or special relativity or
general relativity.  The assumption blocks some avenues of exploration that
open if, in contrast to assuming a global time, we avail ourselves of
freedoms to construct ``local times'' linked to whatever cyclic processes we
choose or invent.  We highlight some of these freedoms in the remarks that
intersperse the next two sections.

\section{Case Studies Involving Entangled Photon Pairs}

We consider two uses of entangled photon pairs.  One use is for quantum key
distribution (QKD); the other use is to synchronize separated clocks.  The
two uses are related, because suitably managed clocks solve what we call the
``sequence-ordering problem'' that arises in quantum key distribution.  In
its most basic form, QKD aims to provide two agents, Alice and Bob, with a
cryptographic key that is theoretically secure against undetected
eavesdropping by Eve \cite{ekert}.  Alice and Bob make use of their key to
communicate privately, that is, to communicate encrypted messages, unreadable
by Eve.

For our first case, reported by Tittel et al. in 2000 \cite{00tittel}, we
discuss a QKD experiment in which a pulsed laser cyclically pumps a
down-converting crystal repetitively with short light pulses, thereby
generating entangled photon pairs, with some efficiency less than 1.  Unlike
cases that follow, this case pumps with a pulsed laser that imposes a
rhythmic cycle of operation on the experiment.  Whether that rhythm is
tightly connected to a standard frequency as defined in the International
System (SI) turns out to be irrelevant.  Another striking feature of this
experiment is the employment of photon pairs entangled in such a way that
phase correlations matter \cite{91franson,90Ou,90RT}. A source $S$ generates
a sequence of weak light pulses, theoretically described as single photons,
into an optical fiber.  The optical fiber forks into two branches, and each
pulse splits into weaker pulses, one propagating along each branch.  After
propagating along the branches, with one branch imposing an extra delay
relative to the other branch, the pulses enter again into a single fiber, one
pulse delayed relative to the other.  In addition, one pulse is offset
relative to the other by an adjustable phase increment $\phi$.  In theory,
what emerges is a single photon consisting of a superposition of an earlier
and a later photon.  The light next passes through a down converter, out of
which comes light explained as a quantum state consisting of a superposition
of tensor products of a pair of ``early photons'' with a pair of ``late
photons.''
\begin{equation}
  |\psi\rangle = \frac{1}{\sqrt{2}}\left(|s\rangle|s\rangle + e^{i\phi}|l\rangle|l\rangle
  \right).
\end{equation}
Here, $|s\rangle$ is the early state and $|l\rangle$ is the later state.  The
light next goes into a fork in the optical fiber, with one branch of the fork
directing the light toward Alice while the other branch directs the light
toward Bob.  As the light reaches Alice it enters another branching and
rejoining with unequal arms, providing an earlier and a later pulse, and with
another controllable phase increment $\alpha$.  The light then goes into
another fork in the light path with each branch of the fork leading to a
sensitive light detector.  Bob operates symmetrically.

The upshot is that Alice and Bob need to operate in a cycle inherited from the pump laser.  This cycle has several distinct phases.
(These phases of a cycle have nothing to do with the phase $\phi$ or $\alpha$
of the narrow-band light pulse.)  Alice has not just one phase of a cycle for
detection, but three distinct phases.  She has phases for early detection,
middle detection, and late detection: early if an early pulse from the source
is registered as traveling the early branch of Alice; middle
for a superposition of an early branch at the source followed by a late
branch at Alice's receiver or vice versa; and late if a late pulse from the
source is registered as traveling the late branch of Alice's receiver.
Ideally, for each cycle, just one of her detectors registers a detection and,
furthermore, registers that detection in just one of the three phases.  Again, Bob
operates the same way.

Although the experiment used a single laboratory clock to drive all three
cycles, that is, the cycles of the source, of Alice, and of Bob, a more basic
consideration is that the source $S$ numbers its pulses and that Alice and
Bob, by estimating transmission delays, number their cycles of reception to
match what is transmitted from the source.  If the source intersperses its
photon pairs with strong light signals, the strong light signals that arrive
at Alice can act as her local clock, and the same for Bob.  This arrangement
allows for Alice and Bob to be mobile, relative to the source and to each
other.  In the mobile extension of the experiment, because of varying
propagation delays, the clocks of Alice, Bob, and the source will tick at
rates that vary relative to, for example, a Global Positioning System time
coordinate.

This experimental design, in which Alice's and Bob's clock ticks are defined by
signals arriving form the source, illustrates a form of synchronization
markedly distinct from that defined by Einstein in special relativity. We
call this condition that meshes a phase of reception with the arrival of a
pulse {\it logical synchronization}.  As discussed in Reference \cite{19MST}, logical
synchronization (but not Einstein synchronization) is possible for two clocks
in relative motion.  To maintain logical synchronization, Bob would take a running average of deviations of arriving pulses from the center of the desired phase and use this running average in a feedback loop to adjust the rate of his clock.\\

\noindent{\bf Remarks:}

\begin{enumerate}
\item Unlike the continuous
wave (CW) operation soon to be discussed, this pulsed operation with its
source-imposed cycle allows for gating of light detectors, thereby reducing
spurious detections \cite{gateQKD}.
\item There is no need for the clock that steps the pulse laser to have any definite relation to a laboratory time standard.  
\end{enumerate}

An alternative method of coordinating Alice's and Bob's detections of photon pairs is the use of time stamps, as will be discussed in the examples below.

\section{Experiments with Continuous-Wave Pumping}
Three more examples come from the laboratory of C. Kurtsiefer at the Centre for Quantum Technologies (CTQ) at the National University of Singapore.
The first two of these three are for QKD, while the third concerns the use of
photon pairs in synchronizing clocks used for other purposes.
As in the example above, all three of the examples from the CTQ involve agents Alice and Bob, and all three have certain additional common features:
\begin{enumerate}
\item  They all involve polarization-entangled photon pairs as short light pulses.
\item The pairs are generated not in the rhythm established by a pulsed
    laser, but by use of a continuous-wave (CW) laser, so that the pair
    production is more or less a Poisson process with unpredictable durations
    between photon pairs.
\item Alice and Bob have separate clocks that they must occasionally adjust.
\item Alice records light detections in a sequence of records and includes a
  reading of her clock in each record of detection.  Bob operates the same
  way.
\item Records made by Alice and Bob are subjected to post-processing in order
  to coordinate comparisons of detections stemming from a common photon pair.
\item The unpredictability of the durations between photon pairs plays an indispensable role.
\end{enumerate}

\subsection{A QKD Experiment in 2006}
The first of the three examples comes from the report of Marcikic et
al.\ \cite{06k}.  In this experiment, Alice employs a CW laser to generate
unpredictably spaced photon pairs.  She records detections and attaches a
reading of her clock to each detection record.  Similarly, Bob records
detections and attaches a reading of his clock to each of his detection
records.  In order to identify which records are to be compared with which,
Bob transmits (classically) his sequence of clock readings to Alice, who then
computes the correlation of her sequence with Bob's sequence.  The
correlation is a function of an offset variable $\tau$ (shown explicitly in
Equations (3) and (4) of Ref. \cite{09k}).  Under the assumption that the offset in
clock readings does not change, (as would be the case if the propagation
delay is constant), the correlation integral has a sharp
peak at a value of $\tau$ that corresponds to the reading of Bob's clock at
the detection of a photon less the reading of Alice's clock at her detection
of a photon from the same entangled pair.  Given this offset, the appropriate
comparisons of paired detections can be made.  For this to work, of course,
Alice's and Bob's clocks have to be both stable enough and close in frequency.

The slowly varying clock offset is used in a feedback loop to limit the drift
of Bob's clock relative to receptions from Alice; that is, the running offset is
fed back to adjust the frequency of Bob's clock relative to Alice's
transmissions.  The same feedback loop actually allows for a more general
operation, in which Bob and Alice can be in gentle relative motion.\\

\noindent{\bf Remarks:}
\begin{enumerate}
\item This method of determining the correspondence between
Alice's and Bob's clock readings depends critically on the unpredictability
of durations between paired-photon emissions. If, as in the preceding
experiment of Tittel et al., the pump laser was periodically pulsed, the
correlation would show periodic peaks and thus be useless for guiding the
adjustment of Bob's clock.
\item The irregular photon pairs can be taken as ticks that mark an
  unpredictable local time.
  \end{enumerate}

\subsection{A QKD Experiment in 2009}
A second experiment from the CTQ was reported by C. Ho
et al., ``Clock synchronization by remote detection of correlated photon
pairs'' \cite{09k}.  The authors report an experiment on quantum key
distribution in which they show how to avoid some otherwise stringent
requirements for hardware synchronization.  In examining their use of clocks,
we speak of Alice and her clock readings $t_{A,i}$ where Ho et al. speak of
reference clock $A$ and its readings $t_{i}$; we also speak of Bob and his
clock readings $t_{B,j}$ where Ho et al. speak of reference clock B and its
readings $t'_j$.  The experiment involves a continuous source $S$ of
entangled photon pairs, with one photon directed to Alice and the other to
Bob.  The advance over the design in the preceding experiment allows the use
of less stable clocks, achieved by an iterative scheme of feedback to adjust not only the offset of Bob's clock readings  relative to Alice's clock readings, but also to steer the relative frequency of the two clocks.  Because of this feedback,
the use of post-processing might be called ``prompt post-processing''; that is, one cannot wait too long before using the post-processed correlation to adjust the clock's relative frequencies.\\

\noindent{\bf Remarks:}
\begin{enumerate}
\item As in the preceding example, the method of using correlations to adjust the
relative rate of Alice's and Bob's clocks allows for Alice and  Bob to be in gentle relative motion.
\item Relative motion of the clocks precludes their satisfying the conditions of
  synchronization specified in special relativity \cite{19MST} yet does not limit the precision with which the clock readings can be made to correspond.
\end{enumerate}

\subsection{An Experiment on Clock Synchronization in 2018}
The third experiment from the laboratory of Kursiefer that we discuss is concerned not with QKD but with clock synchronization per se.  In ``Symmetrical clock synchronization with time-correlated photon pairs'' Lee et al.\ use the correlation technique to ``demonstrate a point-to-point clock synchronization protocol based on bidirectionally exchanging photons produced in spontaneous parametric down conversion (SPDC)'' \cite{18k}.  By employing entangled photons, the authors offer security against some (not all) malicious attacks on synchronization procedures, based on the ability to test violations of Bell inequalities.

The experimental set up doubles that of Marcikic et al.  As in that
experiment, Alice employs a CW laser to generate unpredictably spaced photon
pairs.  She records detections and attaches a reading of her clock to each
detection record.  Similarly, Bob records detections and attaches a reading
of his clock to each of his detection records. For this experiment, however, Bob also
employs a CW laser, so the activity of generation of photon pairs and their
transmission goes on in both directions.  Suppose Alice generates a given
photon pair and detects one photon at reading $t^X_A$ of her clock, and
suppose that Bob detects the other photon of the pair at reading $t^R_B$ of
his clock. (The superscript $X$ is for `transmit,' and $R$ is for `receive'.)
Then $t^R_B-t^X_A$ is, in the notation of Lee et al., $\tau_{AB}$.  Going the
other way around, from Bob to Alice, we have $t^R_A-t^X_B$ is $-\tau_{BA}$
(note the minus sign) in the notation of Lee et al.  Again using ``prompt
post-processing'' to compute timing correlations, and acknowledging that they
``made the strong assumption that the photon propagation times in both
directions were equal,'' the authors show how two correlation peaks combine to generate the needed steering to bring about their form of synchronization.\\

\noindent{\bf Remarks:}
\begin{enumerate}
\item In special relativity synchronization of proper clock $A$ to proper clock $B$, with both clocks fixed relative to some inertial frame, invokes an (idealized) light signal from $A$ at $t_A$ which echoes off $B$ at $t_B$ and returns to $A$ at $t'_A$, so as to satisfy, independent of $t_B$, the equations
  \begin{equation}\label{eq:2}
    t_B=(t'_A+t_A)/2.
  \end{equation}
  An informative discussion of other definitions of  synchronization, applicable to improper clocks and to clocks in a curved spacetime, is given by Perlick in Ref. \cite{perlick07}.
  \item The authors invoke what amounts to (\ref{eq:2}) to express what they mean by ``synchronization.''
\item The authors demonstrate experimentally that their criterion is met to within experimental tolerances for several different lengths of fibers, from which they
  conclude that the design works for clocks in relative motion.  However,
  if by synchronization the authors mean satisfying (\ref{eq:2}), this conclusion
  is not strictly correct because, as pointed out by Perlick, the synchronization relation (\ref{eq:2}) is not symmetric and cannot hold bidirectionally for clocks in relative motion.
  \item We note that although the bidirectional synchronization that accords with (\ref{eq:2}) is ruled out for clocks in relative motion, another form of synchronization is possible, and it is precisely the {\it logical} synchronization required for the bi-directional communications of digital symbols, discussed in Ref. \cite {19MST}.
\end{enumerate}

\section{Discussion}
The habit of thinking that whatever happens must happen in space and time
(or, relativistically, in spacetime) is widespread; nonetheless, we assert
that there is an alternative, for we see spacetime as a mathematical
construct, visible only on the blackboard, for instance in expressions
involving Lorentz transforms.  In spite of legions of experimental evidence
that accord with these blackboard expressions, we distinguish spacetime as a
concept expressed in formulas from a theory of clocks based on relating symbols transmitted to symbols received.

Our proof of the multiplicity of explanations of given evidence, mentioned in
the introduction, led to the later demonstration of unpredictability in the
behavior of physical devices used in experiments \cite{19MST}. We predict
that examples of such unpredictability will be visible in records obtainable
from experiments on entangled light states, such as those above, in the form
of steering commands sent to adjust the clocks of Alice and/or Bob.

Interest in entangled states was advanced by a series of experiments on
violations of Bell inequalities, leading to a particularly clear experiment
on entanglement of polarization in 1982 \cite{aspect}.  An experiment on
entangled states requires a system with several agents working at separated
locations.  Both in setting up an experiment and during its operation, in
principle each agent has occasion to update the quantum state by which that
agent explains the operation of the whole system, that is, the system that
includes other agents.  We like to picture any agent's quantum state as
written into a file local to that agent, on the basis of evidence available
at the moment to the agent.  From this point of view, each agent's quantum
state for the system is ``local'' in the sense of belonging to that agent.
It is to be noted that the quantum state as used by an agent to guide that
agent's local actions cannot sensibly be taken to be a description of reality
as seen by an observer who does not participate in the business of the
system. This view of a quantum state as local to an agent resolves what
otherwise is the serious obstacle to relativistic quantum theory that was
reported by Aharonov and Albert \cite{AA81,AA84b}.

An agent's clock provides the agent with clock readings; that is, Alice's
clock provides her with a ``local time.''  Similarly for Bob, but neither
Alice nor Bob are provided by their clocks with any ``global time
coordinate.''  Thus, in addition to the locality of quantum states, we have
``local times, '' and, as illustrated above, there does not seem to be any
absolute need to invoke any concept of ``global spacetime.'' This point is further
elaborated in Ref. \cite{19MST}.

\section*{Acknowledgment}We are greatly indebted to three reviewers for their trenchant criticisms of an earlier draft.


\begin{thebibliography}{999}

\bibitem{19MST} Myers, J. M.; Madjid, F. H. Synchronization of symbols as
  the construction of times and places. {\em Meas. Sci. Tech.} {\bf 2020}, {\em 31}, 025106. doi:10.1088/1361-6501/ab50dc.

\bibitem{AA81} Aharonov, Y.; Albert, D. Z. Can we make sense out of
  the measurement process in relativistic quantum mechanics? {\em
    Phys. Rev.} {\bf 1981}, {\em D24}, 359--370.

\bibitem{AA84b} Aharonov, Y.; Albert, D. Z. Is the usual notion of time
  evolution adequate for quantum-mechanical systems?  II. Relativistic
  considerations. {\em Phys. Rev.} {\bf 1984}. {\em D29}, 228--234.

\bibitem{zeilinger} Schlosshauer, M.; Kofler, J.; Zeilinger, A.  A
  snapshot of foundational attitudes toward quantum mechanics. {\em arXiv} {\bf 2013}, arXiv:1301.1069.

\bibitem{05aop} Madjid, F.~H.; Myers, J.~M.  Matched detectors
  as definers of force. {\em Ann. Phys.} {\bf 2005},
  {\em 319}, 251--273.
  
\bibitem{00tittel} Tittel, W.; Brendel, J.; Zbinden, H.; Gisin, N. Quantum cryptography using entangled photon in energy-time Bell states.
  {\em Phys. Rev. Lett.}{\bf 2000},
  {\em 84}, 4737--4740.
  
\bibitem{06k} Marcikic, I.;  Lamas-Linares, A.;  Kurtsiefer, C.
Free-space quantum key distribution with entangled photons. {\em
  Appl. Phys. Lett.} {\bf 2006},
{\em 89}, 101122. doi:10.1063/1.2348775.

\bibitem{09k} Ho, C.; Lamas-Linares, A.; Kurtsiefer C. Clock
  synchronization by remote detection of correlated photon pairs. {\em New J. Phys.} {\bf 2009}, {\em 11}, 045011 (13pp).  doi:10.1088/1367-2630/11/4/045011.

\bibitem{18k}
Lee, J.; Shen, L.; Cer\`e, A.;  Troupe, J.;  Lamas-Linares, A.; Kurtsiefer, C. Symmetrical clock synchronization with time-correlated photon pairs. {\em arXiv} {\bf 2018}, arXiv:1812.08450v1.
 
\bibitem{ekert} Ekert, A. K. Quantum cryptography based on Bell's theorem.
  {\em Phys. Rev. Lett.} {\bf 1991},
  {\em 67}, 661--663.


\bibitem{91franson} Franson, J. D.  Two-photon interferometry over large distances. {\em Phys. Rev. A} {\bf 1991},
  {\em 44}, 4552--4555.
  
\bibitem{90Ou} Ou, Z. Y.; Zou, X.Y.; Wang, L.J.; Mandel,  L. Observation of nonlocal interference in separated photon channels. {\em Phys. Rev. Lett.} {\bf 1990}, {\em 66}, 321--324.

\bibitem{90RT} Rarity, J. G.; Tapster, P. R. Experimental violation of
  Bell's inequality based on phase and momentum. {\em Phys. Rev. Lett.} {\bf
    1990}, {\em 64}, 2495--2498.

\bibitem{gateQKD} Brendel, J.; Gisin, N.;  Tittel, W.; Zbinden, H. Pulse energy-time entangled twin-photon source for quantum communication.
  {\em Phys. Rev. Lett.} {\bf 1999},
  {\em 82}, 2594--2597.
  
\bibitem{perlick07}  Perlick, V. On the radar method in
  general-relativistic spacetimes. In  {\em Lasers, Clocks,
  and Drag-Free Control. Exploration of Relativistic Gravity
  in Space}, 2007 ed.;  Springer: Berlin, Germany 2007; pp. 131--152.
  
\bibitem{aspect} Aspect, A.; Dalibard, J.; Roger, G. Experimental test of
  Bell's inequalities using time-varying analyzers. {\em Phys. Rev. Lett.}
  {\bf 1982}, {\em 49}, 1804--1807.


\end{thebibliography}
\end{document}